\documentclass[10pt]{iopart}
\usepackage{graphicx,epsfig}% Include figure files
\usepackage{bm}% bold math
\usepackage{iopams}
\newcommand{\ba}{\begin{eqnarray}}
\newcommand{\ea}{\end{eqnarray}}
\newcommand{\bse}{\numparts}
\newcommand{\ese}{\endnumparts}

\newcommand{\bbq}{\begin{quote}}
\newcommand{\eeq}{\end{quote}}
\graphicspath{{./}{Figuras/}}

\begin{document}

\title[Gravity induced evolution of a magnetized fermion gas with finite temperature]{Gravity induced evolution of a magnetized fermion
gas with finite temperature}

\author{I. Delgado Gaspar$^{\clubsuit}$, A. P\'erez Mart\'\i nez$^{\diamondsuit}$, Roberto A. Sussman$^\spadesuit$ and A. Ulacia
Rey$^{\diamondsuit,\spadesuit}$.}
\address{$^{\clubsuit}$ Instituto de Geof\'isica y Astronom\'ia (IGA). Calle 212 No 2906, La Lisa. La Habana, Cuba. cp-11600.
\\
$^{\diamondsuit}$ Instituto de Cibern\'etica, Matem\'atica, y F\'isica (ICIMAF). Calle 15 No. 309, Vedado. La Habana, Cuba. cp-10400.
\\
$^\spadesuit$ Instituto de Ciencias Nucleares, Universidad Nacional Aut\'onoma de M\'exico (ICN-UNAM). A. P. 70--543, 04510 M\'exico D. F.}
\eads{$^{\clubsuit}$\mailto{idelgado@iga.cu}, $^{\diamondsuit}$\mailto{aurora@icimaf.cu}
, $^\spadesuit$\mailto{sussman@nucleares.unam.mx}, $^{\diamondsuit,\spadesuit}$\mailto{alainulacia@nucleares.unam.mx/alain@icimaf.cu}}

\date{\today}

\begin{abstract} We examine the near collapse dynamics of a self--gravitating magnetized electron gas at finite
 temperature, taken as the source of a Bianchi-I spacetime described by the Kasner metric. The set of Einstein--Maxwell field
equations reduces to a complete and self--consistent system of non--linear autonomous ODE's. By considering a representative
set of initial conditions, the numerical solutions of this system show the gas collapsing into both, isotropic (``point--like'')
and anisotropic (``cigar--like'') singularities, depending on the intensity of the magnetic field. We also examined the behavior
during the collapse stage of all relevant state and kinematic variables: the temperature, the expansion scalar, the magnetic field,
the magnetization and energy density. We notice a significant qualitative difference in the behavior of the gas for a range
of temperatures between the values $\hbox{T}\sim10^{3}\hbox{K}$ and  $\hbox{T}\sim 10^{7}\hbox{K}$.
\end{abstract}

% 98.80.-k Cosmology
% 04.20.-q Classical general relativity
% 04.40.+-b Self-gravitating systems
% 05.30.Fk Fermions systems and electrons
% 95.36.+x Dark Enery
% 95.35.+d Dark Matter

\maketitle
\section{Introduction}

Astrophysical systems provide an ideal scenario to examine the effects of strong magnetic fields
associated with self--gravitating sources under critical conditions. In such conditions, we expect non--trivial coupling between
gravitation and other fundamental interactions (strong, weak and electromagnetic), and from this interplay important
clues of their unification could emerge.

The presence and effects of strong magnetic fields in compact
objects (neutron, hybrids and quark stars) have been studied  in the
literature (see \cite{shapiro,EOS1991,EOS1996,EOS1999,Cardall2000bs,EOS2002,EOS2007,Paulucci:2010uj},
and references quoted therein), assuming various types of equations of state (EOS) has been obtained
and considering in some of these papers numerical solutions of the equilibrium Tolman--Oppenheimer--Volkov (TOV) equation.

As proven in previous work \cite{shapiro,Paulucci:2010uj,Martinez:2003dz,Felipe:2008cm,Ferrer:2010wz},
the presence of a magnetic field is incompatible with spherical symmetry and necessarily introduces
anisotropic pressures. However, by assuming that anisotropies and deviations from spherical symmetry
remain small, several authors \cite{Paulucci:2010uj,Felipe:2008cm,Felipe:2010vr} have managed to compute
observable quantities of idealized static and spherical compact objects under the presence of strong
magnetic fields by means of TOV equations that incorporate these anisotropic pressures. Moreover, it is
evident that much less idealized models would result by considering the magnetic field and its associated
pressure anisotropies in the context of TOV equations under axial symmetric (or at least cylindrically symmetric)
geometries \cite{Paulucci:2010uj}.

As an alternative (though still idealized) approach, and bearing in mind the relation between magnetic fields
and pressure anisotropy, we have examined the dynamics of magnetized self--gravitating Fermi gases as sources
of a Bianchi I space-time \cite{Alain_e-,Alain_2,Alain_n}, as this is the simplest non--stationary geometry that
is fully compatible with a the anisotropy produced by a magnetic field source.

Evidently, a Bianchi I model is a completely inadequate metric for any sort of a compact object, as all geometric
and physical variables depend only on time (and thus it cannot incorporate any coupling of gravity with spatial
gradients of these variables). However, the use of this idealized geometry could still be useful to examine
qualitative features of the local behavior of the magnetized gas under special and approximated conditions.
Specifically, we aim at providing qualitative results that could yield a better understanding of the conditions
approximately prevailing near the center and the rotation axis of less idealized configurations, where the angular
momentum of the vorticity and the spatial gradients of the 4--acceleration and other key variables play a minor dynamical role.

The main objective of the present paper is to include the effect of the temperature in the study that we accomplished
in \cite{Alain_e-}. We aim at addressing the question of whether a finite temperature produces a significant dynamical
effect in ``slowing down'' or reversing the evolution of the magnetized gas in the collapsing regime, as such qualitative
difference may be related to the stability of the self--gravitating configuration. We also analyze the relations between
the magnetization and magnetic field, and the energy density and temperature.

The paper is organized as follow. In section II we derive the EOS for a dense magnetized electron gas at finite temperature.
In section III we lay out the dynamical equations for the evolution of our model by writing up the Einstein--Maxwell
system of equation for the specific source under consideration. Einstein--Maxwell equations are written in section IV
as a system of non--linear autonomous differential equations, rewriting it in section V in terms of physically motivated
dimensionless variables. The numerical analysis of the collapsing solutions and the discussion of the physical results
are given in the section  VI. Our conclusions are presented in sections VII.

\section{Magnetized Fermi gas as a source of a Bianchi I background geometry}

Homogeneous but anisotropic Bianchi I models are described by the Kasner metric

\begin{equation}
ds^{2}=-dt^{2}+Q_{1}\left(t\right)^{2}dx^{2}+Q_{2}\left(t\right)^{2}dy^{2}+Q_{3}\left(t\right)^{2}dz^{2},
\end{equation}
so that spatial curvature vanishes and all quantities depend only on time.
Assuming a comoving frame with coordinates
$x^a=\left[t,x,y,z\right]$ and 4-velocity $u^{a}=\delta_{t}^{a}$, the
energy--momentum tensor for a self--gravitating magnetized gas of free electrons is given by:
\begin{equation}
T^{\,\,a}_{b}=\left(U+P\right)u^{a}u_{b}+P\delta^{\,\,a}_{b}+\Pi^{\,\,a}_{b},
\qquad P=p-\frac{2BM}{3},\label{eq:Tabtensorial}
\end{equation}
where $B$ is the magnetic field (pointing in the $z$ direction), $U$ is the energy density (including the rest energy of the electrons), $M$ is the magnetization of the gas, $P$ is the isotropic pressure and $\Pi^{\,\,a}_{b}$ is the traceless anisotropic pressure tensor:
\begin{equation}\label{eq:Pi_aniso}
\Pi^{\,\,a}_{b}= diag \left[\Pi,\Pi,-2\Pi,0\right],\qquad \Pi=-\frac{BM}{3},
\end{equation}
We can write the energy--momentum (\ref{eq:Tabtensorial}) tensor as:
\begin{equation}
T^{\,\,a}_{b}= \hbox{diag}\left[-U,P_{\bot},P_{\bot},P_{\|}\right],\label{eq:TabMatricial}
\end{equation}
which respectively identifies $P_{\bot}$ and $P_{\|}$ as
the pressure components perpendicular and parallel to the magnetic field.
Notice that the anisotropy in $T^{\,\,a}_{b}$ is produced by the magnetic
field $B$. If this field vanishes, the energy--momentum tensor
reduces to that of a perfect fluid with isotropic
pressure (an ideal gas of electrons complying with Fermi--Dirac statistics).

The equations of
state for this magnetized electron gas can be given in the following form \cite{Can1}:
\begin{eqnarray}
P_{\|}= -\Omega=\lambda\,\beta\,\left[\frac{1}{2}C_{2}\left(\phi,\mu\right)+\sum_{n=1}^{\infty}a{}_{n}^{2}C_{2}\left(\frac{\phi}{a_{n}},
\frac{\mu}{a_{n}}\right)\right]\nonumber
\\
\;\;\;\;\;\;\equiv \lambda\,\Gamma_{\|}\left(\beta,\mu,\phi\right),\label{eq:Pzz}
\\
P_{\bot}=\lambda\,\beta^{2}\sum_{n=1}^{\infty}n C_{1}\left(\frac{\phi}{a_{n}},\frac{\mu}{a_{n}}\right)\equiv\lambda\,
\Gamma_{\bot}\left(\beta,\mu,\phi\right),\label{eq:Pxx}
\\
U=\lambda\,\beta\left[\frac{1}{2}C_{3}\left(\phi,\mu\right)+\sum_{n=1}^{\infty}a{}_{n}^{2}C_{3}\left(\frac{\phi}{a_{n}},
\frac{\mu}{a_{n}}\right)\right]\equiv\lambda\,\Gamma_{U}\left(\beta,\mu,\phi\right).
\end{eqnarray}
The particle number density is,
\begin{equation}
 \fl
\eta=\frac{\lambda}{m_{e}}\,\beta\left[\frac{1}{2}C_{4}\left(\phi,\mu\right)+\sum_{n=1}^{\infty}a_{n}C_{4}\left(\frac{\phi}{a_{n}},
\frac{\mu}{a_{n}}\right)\right]\equiv\frac{\lambda}{m_{e}}\,\Gamma_{\eta}\left(\beta,\mu,\phi\right),\label{eq:Es etha}
\end{equation}
with $a_{n}=\sqrt{1+2n\beta}$ and $C_1,\,C_2,\,C_3,\,C_4$ are given by:
\begin{eqnarray}
%a_{n}=\sqrt{1+2n\beta}\label{eq:an}
%\\
C_{1}\left(\phi,\mu\right)=\intop_{0}^{\infty}\frac{1}{1+\exp\left(\frac{\sqrt{1+x^{2}}-\mu}{\phi}\right)}\,
\frac{dx}{\sqrt{1+x^{2}}},\label{eq:C1}
\\
C_{2}\left(\phi,\mu\right)=\intop_{0}^{\infty}\frac{x^{2}}{1+\exp\left(\frac{\sqrt{1+x^{2}}-\mu}{\phi}\right)}\,
\frac{dx}{\sqrt{1+x^{2}}},
\\
C_{3}\left(\phi,\mu\right)=\intop_{0}^{\infty}\frac{\sqrt{1+x^{2}}}{1+
\exp\left(\frac{\sqrt{1+x^{2}}-\mu}{\phi}\right)}dx,
\\
C_{4}\left(\phi,\mu\right)=\intop_{0}^{\infty}\frac{1}{1+\exp\left(\frac{\sqrt{1+x^{2}}-\mu}{\phi}\right)}dx.\label{eq:C4}
\end{eqnarray}
The equations (\ref{eq:Pzz}) and (\ref{eq:Pxx}) can be accommodated as \cite{Martinez:2003dz,Chaichian:1999gd}:
\begin{equation}
P_{\bot}=P_{\|}-B\emph{M},\label{eq:Pxx-Pzz}
\end{equation}
where $M$, the magnetization, can be written in the following form:
\begin{equation}
M=M_{0}\Gamma_{\emph{M}}\left(\beta,\mu,\phi\right),
\end{equation}
with $\emph{M}_{0}=2\pi\mu_{B}/(\lambda_{c}^{3})$ and $\Gamma_{\emph{M}}$ is obtained of the expressions (\ref{eq:Pzz}), (\ref{eq:Pxx}) and (\ref{eq:Pxx-Pzz}).

In the previous expressions $\mu$ is the dimensionless chemical potential
normalized by the rest energy, $\phi=kT/m_{e}$, $\beta=B/B_{c}$,
where $B_{c}=4.414 \times 10^{13}$ G is the critical magnetic field, $\mu_{B}$ is the Bohr magneton and
$\lambda=8\pi m_{e}/(\lambda_{c}^{3})$ where $\lambda_{c}$ is the Compton wavelength of the electron.

In Appendix A the integrals (\ref{eq:C1})--(\ref{eq:C4}) have been transformed into equivalent expressions in order to facilitate the numerical calculations.

\section{Einstein--Maxwell equations}

Since we are interested in the critical relativistic regimen, the dynamics of the magnetized gas whose EOS we have described in the previous section must be studied through the Einstein field equations in the framework of General Relativity:
\begin{equation}
G_{\mu\nu}=R_{\mu\nu}-\frac{1}{2}Rg_{\mu\nu}=\kappa T_{\mu\nu},\label{eq:Ecuac Einstein}
\end{equation}
together with the balance equations of the energy--momentum tensor and Maxwell's equations,
\begin{eqnarray}
T^{\mu\nu}\,_{;\nu}=0,\label{eq:conseracion E-P}
\\
%\begin{alignedat}{1}
F^{\mu\nu}\,_{;\nu}=0,  \qquad F_{\left[\mu\nu;\alpha\right]}=0,
%\end{alignedat}
\label{eq:ecuac d Maxwell}
\end{eqnarray}
where $\kappa=8\pi G_{N}$ and $G_{N}$ is Newton's gravitational
constant, while square brackets denote anti-symmetrization in (\ref{eq:ecuac d Maxwell}).

Assuming absence of annihilation/creation processes, so that particle numbers are conserved, leads to the following conservation equation:
\begin{equation}
n^{\alpha}\,_{;\alpha}=0,\qquad n^{\alpha}=\eta \,u^{\alpha},\label{eq:ecuac conservacion No. d particulas}
\end{equation}
where $\eta$ es the particle number density. From the field equations (\ref{eq:Ecuac Einstein}) we obtain:
\begin{eqnarray}
-G^{\,\,x}_{x}=\frac{\dot{Q_{2}}\dot{Q_{3}}}{Q_{2}Q_{3}}+\frac{\ddot{Q_{2}}}{Q_{2}}+\frac{\ddot{Q_{3}}}{Q_{3}}=-\kappa P_{\bot},\label{eq:Gxx}
\\
-G^{\,\,y}_{y}=\frac{\dot{Q_{1}}\dot{Q_{3}}}{Q_{1}Q_{3}}+\frac{\ddot{Q_{1}}}{Q_{1}}+\frac{\ddot{Q_{3}}}{Q_{3}}=-\kappa P_{\bot},\label{eq:Gyy}
\\
-G^{\,\,z}_{z}=\frac{\dot{Q_{1}}\dot{Q_{2}}}{Q_{1}Q_{2}}+\frac{\ddot{Q_{1}}}{Q_{1}}+\frac{\ddot{Q_{2}}}{Q_{2}}=-\kappa P_{\|},\label{eq:Gzz}
\\
-G^{\,\,t}_{t}=\frac{\dot{Q_{1}}\dot{Q_{2}}}{Q_{1}Q_{2}}+\frac{\dot{Q_{1}}\dot{Q_{3}}}{Q_{1}Q_{3}}+
\frac{\dot{Q_{2}}\dot{Q_{3}}}{Q_{2}Q_{3}}=\kappa U.\label{eq:Gtt}
\end{eqnarray}
where $\dot{Q}=Q_{;\alpha}u^{\alpha}=Q_{,t}$. From the conservation
of the energy--momentum tensor (\ref{eq:conseracion E-P}) we obtain:
\begin{equation}
\dot{U}=-\left(\frac{\dot{Q_{1}}}{Q_{1}}+\frac{\dot{Q_{2}}}{Q_{2}}\right)\left(P_{\bot}+U\right)-
\frac{\dot{Q_{3}}}{Q_{3}}\left(P_{\|}+U\right).
\end{equation}
Maxwell's equations (\ref{eq:ecuac d Maxwell}) yield:
\begin{equation}
\frac{\dot{Q_{1}}}{Q_{1}}+\frac{\dot{Q_{2}}}{Q_{2}}+\frac{1}{2}\frac{\dot{B}}{B}=0,
\end{equation}
and from the particle number conservation (\ref{eq:ecuac conservacion No. d particulas}) leads to:
\begin{equation}
\dot{\eta}+\left(\frac{\dot{Q_{1}}}{Q_{1}}+\frac{\dot{Q_{2}}}{Q_{2}}+\frac{\dot{Q_{3}}}{Q_{3}}\right)\eta=0.\label{eq:eta}
\end{equation}

\section{Local kinematic variables}

Einstein--Maxwell field equations are second order system of ordinary differential equations (ODE's). In order to work with a first order system of ODE's, it is useful and convenient to rewrite these equations in terms of covariant kinematic variables that convey the geometric effects on the kinematics of local fluid elements through the covariant derivatives of $u^\alpha$. For a Kasner metric in the comoving frame endowed with a normal geodesic 4--velocity, the only non--vanishing kinematic parameters are the expansion scalar, $\Theta$, and the shear tensor $\sigma_{\alpha\beta}$:
\begin{equation} \Theta=u^{\alpha}\,_{;\alpha}\,,\qquad \sigma_{\alpha\beta}=u_{(\alpha;\beta)}-\frac{\Theta}{3}h_{\alpha\beta}\,,\end{equation}
where $h_{\alpha\beta}=u_{\alpha}u_{\beta}+g_{\alpha\beta}$ is the projection tensor and rounded brackets denote symmetrization. These parameters take the form:
\begin{eqnarray}
\Theta=\frac{\dot{Q_{1}}}{Q_{1}}+\frac{\dot{Q_{2}}}{Q_{2}}+\frac{\dot{Q_{3}}}{Q_{3}}\,,\label{eq:def theta}
\\
\sigma^{\,\,\alpha}_{\beta}=\hbox{diag}\,\left[\sigma^{\,\,x}_{x},\sigma^{\,\,y}_{y},\sigma^{\,\,z}_{z},0\right]=\hbox{diag}\,\left[\Sigma_{1},\Sigma_{2},\Sigma_{3},0\right]\,,\label{eq:def sigma}
\end{eqnarray}
where:
\begin{eqnarray}
%\begin{aligned}
\Sigma_{a}=\frac{2}{3}\frac{\dot{Q_{a}}}{Q_{a}}-\frac{1}{3}\frac{\dot{Q_{b}}}{Q_{b}}-
\frac{1}{3}\frac{\dot{Q_{c}}}{Q_{c}},\qquad & a\neq b\neq c\,\left(a,b,c=1,2,3\right).
%\end{aligned}
\label{eq: componentes}
\end{eqnarray}
The geometric interpretation of these parameters is straightforward:  $\Theta$
represents the isotropic rate of change of the 3--volume of a fluid element, while
$\sigma^{\,\,\alpha}_{\beta}$ describes its rate of local deformation along different spatial directions given by its eigenvectors.
Since the shear tensor is traceless: $\sigma^{\,\,\alpha}_{\alpha}=0$,
it is always possible to eliminate any one of the three quantities $\left(\Sigma_{1},\Sigma_{2},\Sigma_{3}\right)$ in terms of the other two. We choose to eliminate $\Sigma_{1}$
as a function of $\left(\Sigma_{2},\Sigma_{3}\right)$.
By using equations (\ref{eq:def theta}) and (\ref{eq: componentes})
we can re--write the second derivatives of the metric functions
in (\ref{eq:Gxx}), (\ref{eq:Gyy}) y (\ref{eq:Gzz}) as first order
derivatives of $\Theta$, $\Sigma_{2}$ and $\Sigma_{3}$. After some algebraic manipulations it is possible to transform equations (\ref{eq:Gxx})-(\ref{eq:eta}) as a first order system of autonomous ODE's:
%
%\begin{subequations}
\bse
\begin{eqnarray}
\dot{\Sigma}_{2}=\frac{\varkappa}{3}\left(P_{\bot}-P_{\|}\right)-\Theta\Sigma_{2},\label{eq:sistema1}
\\
\dot{\Sigma}_{3}=\frac{2\varkappa}{3}\left(P_{\|}-P_{\bot}\right)-\Theta\Sigma_{3},\label{eq:sistema2}
\\
\dot{B}=2B\left(\Sigma_{3}-\frac{2}{3}\Theta\right),\label{eq:sistema3}
\\
\dot{\mu}=\frac{1}{\hbox{Det}}\left[ f_{1}\eta_{,T}-f_{2}U_{,T}\right],\label{eq:sistema4}
\\
\dot{T}=\frac{1}{\hbox{Det}}\left[ f_{2}U_{,\mu}-f_{1}\eta_{,\mu}\right],\label{eq:sistema5}
\end{eqnarray}
\ese
%\end{subequations}
%
together with the following constraint:
\begin{equation}
\varkappa U=-\left(\Sigma_{2}\right)^{2}-\Sigma_{2}\Sigma_{3}-\left(\Sigma_{3}\right)^{2}+\frac{\Theta^{2}}{3},\label{eq: vinculoSAdi}
\end{equation}
where:
\begin{eqnarray}
 \hbox{Det} &\equiv& U_{,\mu}\eta_{,T}-U_{,T}\eta_{,\mu},
\\
 f_{1} &=& \left(\Sigma_{3}-\frac{2\Theta}{3}\right)\left(P_{\bot}+U\right) -\left(\Sigma_{3}+\frac{\Theta}{3}\right)\left(P_{\|}+U\right)-U_{,B}\dot{B},
\\
 f_{2} &=& \Theta\eta+\eta_{,B}\dot{B}.
\end{eqnarray}
These first order equations form a complete and self--consistent system
whose numeric integration fully determines
$\Sigma_{2}$, $\Sigma_{3}$, $B$, $\mu$, $T$, and thus allows us to study
the dynamical evolution of a local volume element of a gas of magnetized electrons.
\section{Dynamical equations}
By introducing the following dimensionless evolution parameter,
\begin{eqnarray}
%\begin{aligned}
H=\frac{\Theta}{3},\qquad & \frac{d}{d\tau}=\frac{1}{H_{0}}\frac{d}{dt},
%\end{aligned}
\label{eq:func1 adimen}
\end{eqnarray}
together with the dimensionless variables,
\begin{eqnarray}
%\begin{alignedat}{2}
\fl
 {\mathcal{H}}=\frac{H}{H_{0}},\qquad  S_{2}=\frac{\Sigma_{2}}{H_{0}},\qquad  S_{3}=\frac{\Sigma_{3}}{H_{0}},\qquad  \beta=\frac{B}{B_{c}},\qquad\mu=\frac{\tilde\mu}{m_{e}},
%\end{alignedat}
\label{eq:func adimen}
\end{eqnarray}
where we have denoted by $\tilde\mu$ the usual chemical potential and $H_{0}$ is a constant with inverse length units that sets the
characteristic length scale of the system, which we have chosen as
$3H_{0}^{2}=\kappa\lambda\Rightarrow H_{0}=0.86 \times 10^{-10}\,\hbox{m}^{-1}$, so
that $1/H_{0}\cong 1.15\times 10^{10}\, \hbox{m}$ is of the order of magnitude of
an astronomic unit. It indicates that our simplified model is examined on local scales smaller than cosmic scales.  In cosmological sources and models \cite{MTWGravitation} $H_{0}=0.59 \times 10^{-26}\,\hbox{m}^{-1}$ would play the role of the Hubble scale constant, this value is a much greater length scale. The
functions $S_{2}$ and $S_{3}$ are the components of the shear tensor
normalized with this scale, while $\tau$ is the dimensionless time.
Substituting (\ref{eq:func adimen}) into the system (\ref{eq:sistema1})--(\ref{eq:sistema5})
we obtain:
\bse
\begin{eqnarray}
S_{2,\tau}=\Gamma_{\perp}-\Gamma_{\|}-3\mathcal{H}S_{2},\label{eq:sistadimen-a}
\\
S_{3,\tau}=2\left(\Gamma_{\|}-\Gamma_{\perp}\right)-3\mathcal{H}S_{3},
\\
\beta_{,\tau}=2\beta\left(S_{3}-2\mathcal{H}\right),
\\
\tilde{\mu}_{,\tau}=\frac{1}{\hbox{Det}}\left(\Gamma_{\eta,\phi}\widetilde{f}_{1}+\Gamma_{U,\phi}\widetilde{f}_{2}\right),
\\
\phi_{,\tau}=-\frac{1}{\hbox{Det}}\left(\Gamma_{\eta,\mu}\widetilde{f}_{1}+\Gamma_{U,\mu}\widetilde{f}_{2}\right).\label{eq:sistadimen-e}
\end{eqnarray}
\ese
while the constraint (\ref{eq: vinculoSAdi}) becomes,
\begin{equation}
3\Gamma_{U} = -S{}_{2}^{2}-S{}_{3}^{2}-S_{2}S_{3}+3\mathcal{H}^{2},\label{eq:vincadimen}
\end{equation}
and the auxiliary parameters of the previous system take the form:
\bse
\begin{eqnarray}
\widetilde{f}_{1}=\left(S_{3}-2\mathcal{H}\right)\left(\Gamma_{\perp}-2\Gamma_{U,\beta}\beta\right)-
\left(S_{3}+\mathcal{H}\right)\Gamma_{\parallel}-3\Gamma_{U}\mathcal{H},
\\
\widetilde{f}_{2}=3\mathcal{H}\Gamma_{\eta}+2\Gamma_{\eta,\beta}\left(S_{3}-2\mathcal{H}\right),
\\
\hbox{Det}=\Gamma_{U,\mu}\Gamma_{\eta,\phi}-\Gamma_{U,\phi}\Gamma_{\eta,\mu}.
\end{eqnarray}
\ese
In the following section we undertake the numerical study of this system, focusing concretely in the the collapsing regime.
\section{Numeric analysis and physical interpretation}
\begin{figure}[htbp]\label{los2}
% \centering
\begin{center}
\includegraphics[width=10cm]{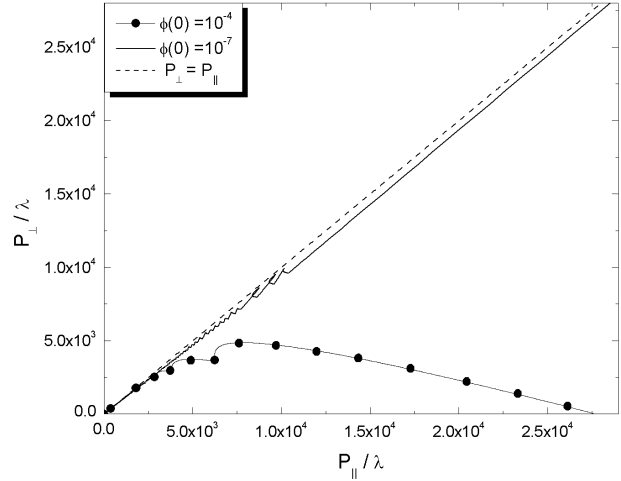}
\caption{\label{fig:G0} Pressure parallel and perpendicular to the
magnetic field. We plot $P_{||}/\lambda$ vs $P_{\perp}/\lambda$ for
initial temperatures $\phi(0)=10^{-7},\,10^{-4}$ where
$\lambda=8\pi m_{e}/(\lambda_{c}^{3})$. } The case $\beta=0$ is
shown for reference.
\end{center}
\end{figure}

%
%
%%%% Rollo sobre las grafica de las presiones
%
Since the energy--momentum tensor (\ref{eq:TabMatricial}) takes the perfect fluid form ($P_{||}=P_\perp$) for
zero magnetic field, we can identify the magnetic field as the factor introducing anisotropy in the dynamical behavior
of the fermionic gas. In particular, this anisotropy in the stress ($P_{||}\ne P_\perp$) must yield different
evolution in different directions, which must be evident in a critical stage such as the collapsing regime.
Intuitively, we expect an isotropic point--like singularity if the pressure is isotropic, as pressure diverges in
all direction, but a large pressure anisotropy (which necessarily corresponds to large magnetic field) should
lead to a qualitatively different direction dependent critical behavior of the pressure that should result in an
anisotropic cigar--like singularity characterized by the divergence of only the pressure parallel to the magnetic
field. As we show in figure \ref{fig:G0}, the pressure parallel and perpendicular for different initial temperatures
does exhibit the expected behavior: for initial conditions of small temperatures $\phi(0)=10^{-7}$ ($\sim 10^3$ K) we have
$P_{||}\approx P_\perp$ and a point singularity: $P_{||},\,P_\perp\to\infty$, but for a larger initial value
$\phi(0)=10^{-4}$ ($\sim 10^6$ K) we have the highly anisotropic evolution $P_{||}\to\infty$ with $P_\perp\to 0$
that signals a cigar--like singularity (the case $P_{||}= P_\perp$ with zero magnetic field is shown as
a comparison).

Since $\mathit{V}=\sqrt{-\det g_{\alpha\beta}}=Q_{1}Q_{2}Q_{3}$,
we obtain by means of (\ref{eq:def theta}) and (\ref{eq:func1 adimen}) the local volume in terms of ${\mathcal{H}}$:
\begin{equation}
\mathit{V}\left(\tau\right)=\mathit{V}\left(0\right)\exp\left(3\intop_{\tau=0}^{\tau}\mathcal{H}d\tau\right),\label{eq: V(tau)}
\end{equation}
where we remark that the sign of $\mathcal{H\left(\tau\right)}$ implies expansion
if ${\mathcal{H\left(\tau\right)}}>0$, and collapse if ${\mathcal{H\left(\tau\right)}}<0$. Besides
this point, equations (\ref{eq:def theta}) and (\ref{eq: componentes}) lead to:
\begin{equation}
Q_{i}(\tau)=Q_{i}(0)\exp\left[\,\,\intop_{\tau=0}^{\tau}\left(\mathcal{H}+S_{i}\right)d\tau\right],\qquad\qquad\left(i=1,2,3\right).\label{eq: Q(tau)}
\end{equation}
where $S_{1}=-\left(S_{2}+S_{3}\right)$.

For the numerical study we assume a magnetized electron gas at high density: $\mu_e(0)=2$, which
means that the chemical potential is $2m_{e}$. The initial values of the magnetic field and
temperature were chosen in the ranges  $\beta(0)\sim10^{-5}$ to $\beta(0)\sim 10^{-4}$ and $\phi\sim10^{-7}$ to $\phi\sim 10^{-3}$
respectively
\footnote{Let us remark that  $\phi=kT/m_{e}$, $\beta=B/B_{c}$,
with $B_{c}=4.414\times10^{13}\hbox{G}$ and  $\mu$ is the chemical potential normalized $\mu_e=\mu_e/m_{e}$, the magnetic
field is in the range $B\thicksim 10^{8}\hbox{G}$ to $B\thicksim 10^{9}\hbox{G}$ and the temperature varies between
$T\sim10^{3}\hbox{K}$ and $T\sim 10^{7}\hbox{K}$.}
 Together with $\mathcal{H}\left(0\right)<0$, we consider the conditions $S_{2}(0)=0$, $S_{3}(0)=0,+1$, which
 correspond to the cases with zero initial deformation and initial deformation (shear) in the direction of $z$
 axis respectively. The calculation has been done using the fourth-order Runge--Kutta method with the local
 truncation relative error less than $10^{-6}$.

\begin{figure}[htbp]
\centering
\begin{center}
\includegraphics[width=10cm]{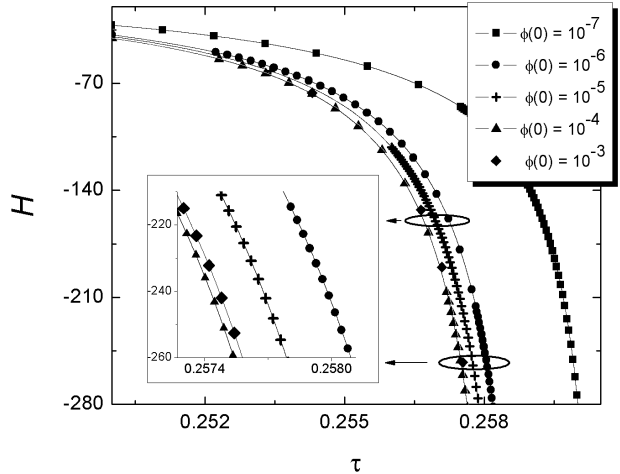}
\caption{\label{fig:G1} Numerical solutions for the expansion scalar $\mathcal{H}\left(\tau\right)$. The
 box in the lower corner is an amplification of the graphic.}
\end{center}
\end{figure}

The numerical solutions for the function $\mathcal{H}$ for the assumed values $\mathcal{H}(0)$ show that
$\mathcal{H}\rightarrow-\infty$, which implies that the volume element evolves to a singularity (see
equation (\ref{eq: V(tau)})). This is exemplified in figure \ref{fig:G1}, where numerical
solutions are displayed for the expansion scalar. These curves correspond to different values of the
initial temperature in the range $\phi(0)=10^{-7}$ to $\phi(0)=10^{-3}$, fixing the rest of the initial
conditions on the values $\mu(0)=2$, $\beta(0)=5\times10^{-5}$, $S_{2}(0)=0$ and $S_{3}(0)=1$. Notice that
in this regime (as given by these initial conditions) where the high densities are dominant, we do not obtain
a direct relation between the values of initial temperatures and the collapse time.
In all the configurations the strength of the magnetic field  diverges during the collapse
of the volume element: this happens independently of the initial conditions.
However, the collapsing time diminishes when the values of the initial magnetic field increases, which
agrees with the results obtained in \cite{Alain_e-}, where $T=0$ was assumed.

The expression (\ref{eq: Q(tau)}) shows how the evolution of the terms $S_{i}+\mathcal{H}$ to $\pm\infty$ implies that the metric
coefficient  $Q_{i}$ evolves to either $+\infty$ or $0$. This evolution is characteristic of an anisotropic
``cigar-like'' singularity, since two  metric coefficients  evolve to zero and the third one to $+\infty$, whereas when all
metric coefficients tend to zero, the singularity is isotropic ``point--like''. See the definitions of these types of
singularities in \cite{dtipo_d_singul}.

For initial conditions with zero shear: $S_{1}(0)=S_{2}(0)=S_{3}(0)=0$, we always
obtain a point--like singularity, independently of the selected values for the remaining
initial conditions. However, when the initial deformation is positive in the direction of the magnetic field ($z$ axis),
we can obtain either cigar--like or point--like singularities.
\begin{figure}[htbp]
\centering
\begin{center}
\includegraphics[width=10cm]{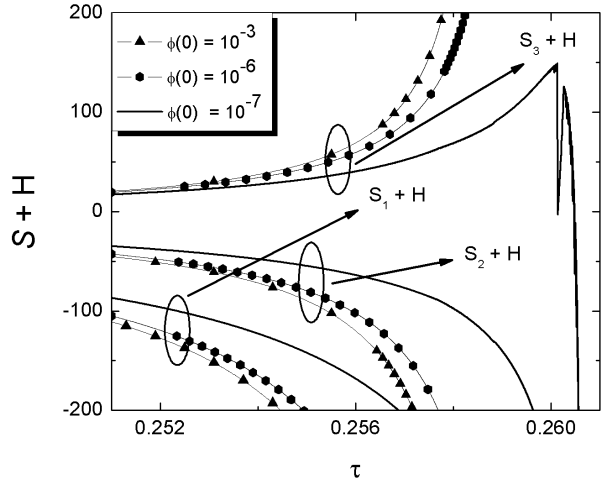}
 \caption{\label{fig:G2} The plots of the functions $S_{i}+\mathcal{H}$ for $i=1,2,3$. The
system has an initial deformation (shear) in the same direction of the
magnetic field along the $z$ axis. The collapse is in the form of
a ``cigar--like'' singularity in the $z$ direction for
$\phi(0)=10^{-3},10^{-6}$ and a ``point--like'' singularity for
$\phi(0)=10^{-7}$.}
\end{center}
\end{figure}
The nine curves displayed in figure \ref{fig:G2} correspond to three sets
of functions: $S_{i}+\mathcal{H}\;\;\left(i=1,2,3\right)$, each one corresponding to these different values of
the temperature: $\phi\left(0\right)=10^{-7},\,10^{-6},\,10^{-3}$. All curves were obtained taking as initial
conditions: $\mu(0)=2$, $\beta(0)=5\times10^{-5}$, $S_{2}\left(0\right)=0$ and $S_{3}\left(0\right)=1$. For
the higher initial values of temperature $\phi\left(0\right)=10^{-6},\,10^{-3}$ the gas collapses into a cigar--like
singularity in the direction of the magnetic field, whereas for the lower value $\phi\left(0\right)=10^{-7}$ we obtain a
point-like singularity. As expected, an anisotropic singularity occurs as the magnetic field increases, even for
relatively low initial temperatures if the initial magnetic field $\beta(0)$ is sufficiently large.

For initial values in the range $\phi(0)\sim10^{-7}$ to $\phi(0) \sim 10^{-4}$ the temperature decreases
%                                                   CORRECCION ALAIN 
%   SE VE CLARAMENTE EN LA GRAFICA de la FIG 6, QUE EL TEMPERATURA, DESPUES DE DISMINUIR, AUMENTA SOBRE VALORES CERCANOS AL 
%        TIEMPO DE COLAPSO por eso puse la oracion "although grows up quickly for values near the collapse time ``
%
as the collapse proceeds, although grows up quickly for values near the collapse time, while for $\phi(0)\sim10^{-3}$
and higher values it increases.

\begin{figure}[htbp]
\centering
\begin{center}
\includegraphics[width=10cm]{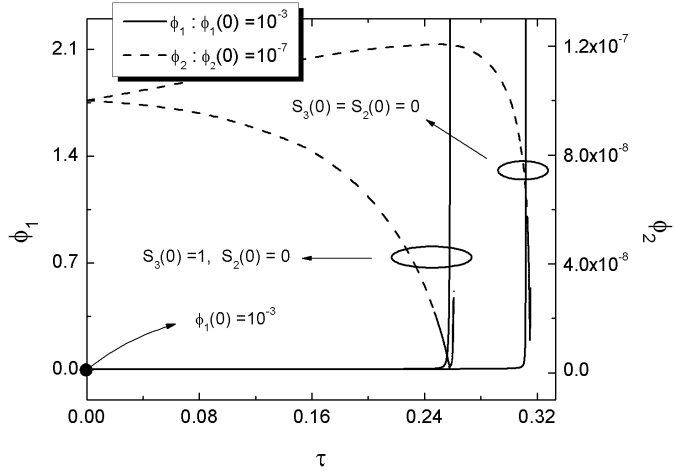}
 \caption{\label{fig:G3} Temperature of the system. The left hand side of the graph displays the
 temperature values ($\phi_{1}$) corresponding to the two solid curves. The right hand side displays the
 temperature values ($\phi_{2}$) corresponding to the two dashed curves.}
\end{center}
\end{figure}

We display in figure \ref{fig:G3} two dashed dashed curves for the evolution of the temperature from
the initial value $\phi\left(0\right)=10^{-3}$, together with solid curves that depict temperatures
starting at $\phi\left(0\right)=10^{-7}$. In both cases the initial conditions correspond to zero
initial shear and positive deformation in the $z$ axis.
The initial values of magnetic field and chemical potential were fixed at $\beta(0)=5 \times 10^{-5}$ and $\mu(0)=2$
respectively. Notice that the symmetric configuration delays the collapse (yields a longer collapsing time), which may
indicate a connection with stability of local fluid elements in compact objects. This could provide an important clue on
the stability of a compact object made of a dense magnetized gas. However, verifying this possibility is beyond the
scope of this article.

\begin{figure}[htbp]
\centering
\begin{center}
\includegraphics[width=10cm]{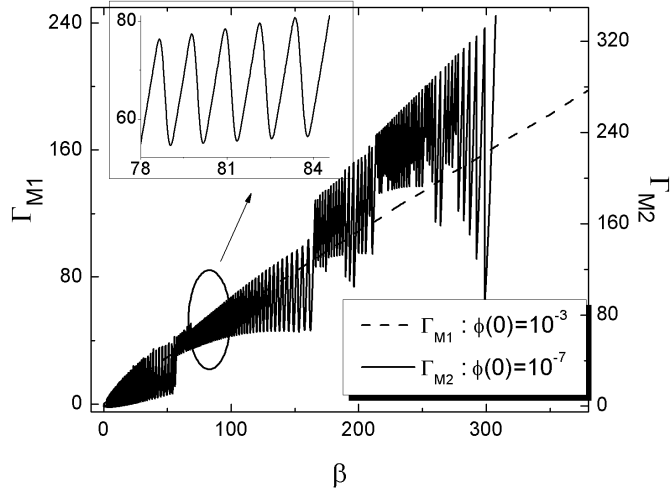}
 \caption{\label{fig:G4} Magnetization  versus magnetic field. The axis of the right hand side
($\Gamma_{M2}$) depicts the magnetization (the solid curve), while the axis of the left hand side ($\Gamma_{M1}$)
corresponds to the dashed curve.}
\end{center}
\end{figure}

\begin{figure}[htbp]
\centering
\begin{center}
\includegraphics[width=10cm]{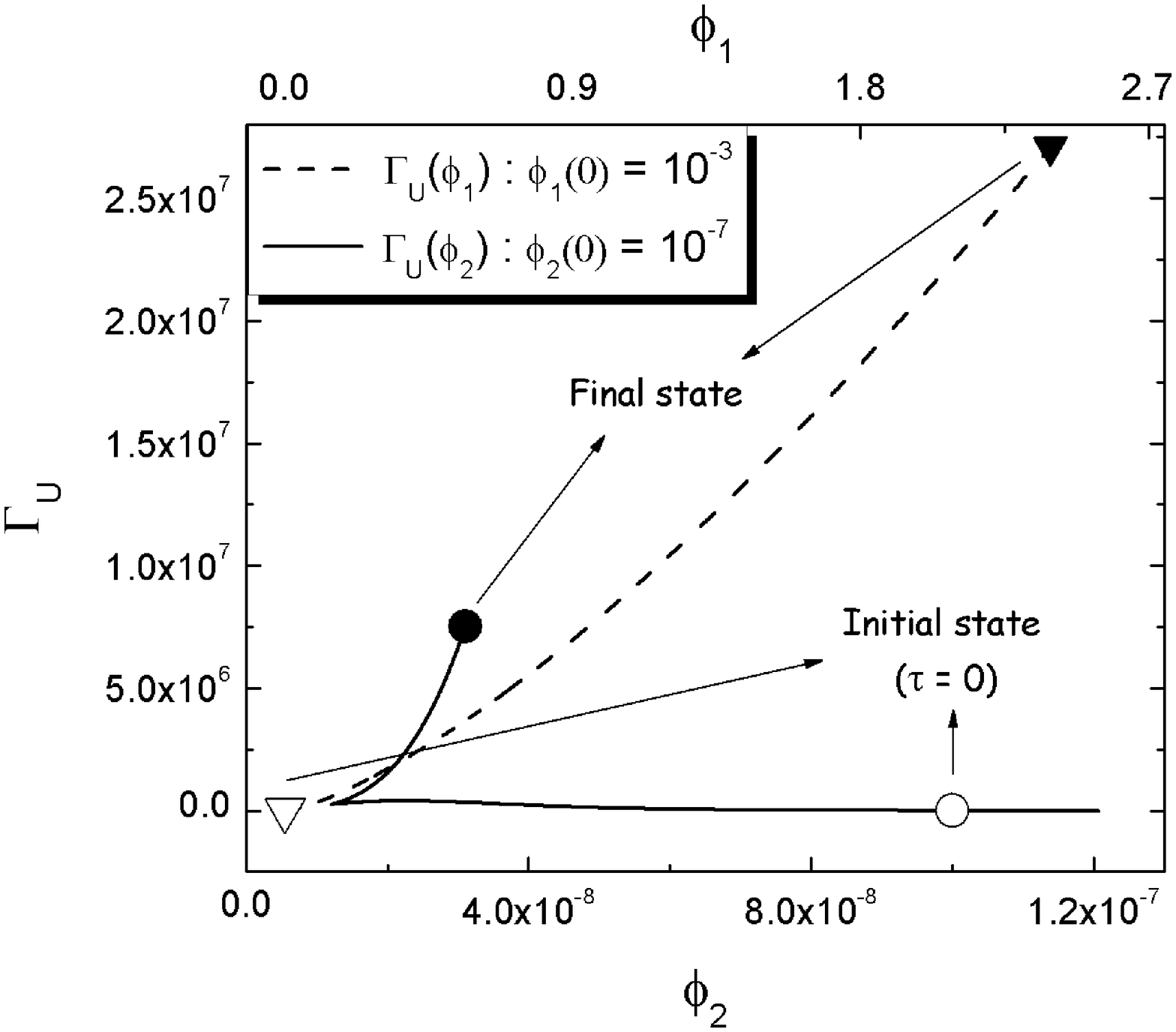}
 \caption{\label{fig:G5}Energy density versus temperature. The temperature values of the curve
draws with solid line should be observed in the lower axis ($\phi_{2}$) and
in the top axis ($\phi_{1}$), those of the curve drawn with dashed line.}
\end{center}
\end{figure}

We examine now the behavior of the magnetization and the energy density of the system during the
collapse. We remark that the $\tau$--depending functions:
$\beta,\; \mu,\; \phi$ follow from the numerical solution of the system
(\ref{eq:sistadimen-a})--(\ref{eq:sistadimen-e}) and
(\ref{eq:vincadimen}), which allows us to compute the magnetization
$M\left(\beta, \;\mu, \;\phi\right)$ and $U\left(\beta, \;\mu,
\;\phi\right)$ for all values of $\tau$. We have depicted
curves of  $M$
vs. the magnetic field and the energy density
vs. temperature
\footnote{Remember that $M$ and $U$ can be
written in the following form: $M = M_{0}\Gamma_{M}$ and
$U=\lambda\Gamma_{U}$.}
for initial conditions: $\beta\left(0\right)=5\times10^{-5}$,
$S_{2}\left(0\right)=S_{3}\left(0\right)=0$, and for two different values
of the initial temperature $\phi\left(0\right)=10^{-7},10^{-3}$. These curves
are shown in figures \ref{fig:G4} and \ref{fig:G5}.

Notice how at low temperature values the magnetization has an
oscillatory behavior with respect to the magnetic field, which
correspond to the well known Haas--van Alphen effect \cite{Haas-van
Alphen}. In contrast, for high initial temperatures, the system is
heated as the collapse proceeds, with the magnetization
behaving monotonically with respect to the magnetic field. These results agree
with the study performed in \cite{Oliver}, where the authors
expressed the magnetization of an electron gas at finite temperature
as the sum of two terms: one oscillatory and the other monotonic, which predominate at
respectively low and high temperatures.

On the other hand, the energy density increases with the
temperature for the initial value $\phi\left(0\right)=10^{-3}$, which does not occurs for the initial value
$\phi\left(0\right)=10^{-7}$.

The behavior of the magnetization and the energy density can be
explained by taking into account that for the initial value
$\phi\left(0\right)=10^{-7}$, as we discussed above, the system is
cooled (temperature decreases) during the collapse process with an increasing magnetic field.
This is an indication that the effects of the magnetic field (alignment of electron spins
with the magnetic field)
predominate over the effects of temperature (random motion) in the system, a situation
in which the Haas--van Alphen oscillations appear in the magnetization curves, with energy
density decreasing with temperature in the collapse process. On the other hand,
for high initial temperatures ($\phi\left(0\right)=10^{-3}$), the
system is heated (temperature increases) during the collapse, which shows
the predominance of the effect of
temperature over the effect of the field, and showing the
magnetization and the energy density following a monotonic behavior with
respect to the magnetic field and the temperature, respectively.
That is to say, when $\phi\left(0\right)\sim10^{-7}$ the temperature
of the system decreases in the collapse process, while the magnetic field increases and the
gas evolves to strong magnetic field regimen, where the effect of the
temperature is negligible in a first approximation. This is not the case for
$\phi\left(0\right)\sim10^{-3}$, for which  the temperature rises
significantly during the collapse and its effects are significant.

\begin{figure}[htbp]
\centering
\begin{center}
\includegraphics[width=10cm]{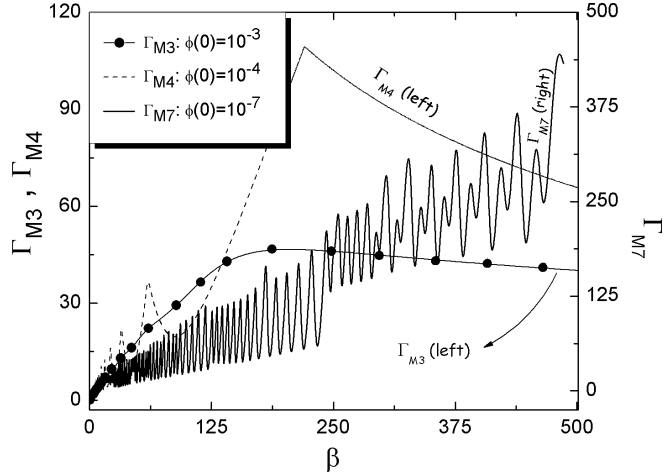}
\caption{\label{fig: disc1} The figure displays the plot of magnetization $M$ for initial temperatures
$\phi(0)=10^{-3},\,10^{-4}, \,10^{-7}$ and $\Gamma_{M3}$, $\Gamma_{M4}$ (vertical axis in the left hand side)
and $\Gamma_{M7}$ (vertical axis in the right hand side) vs magnetic field $\beta$. We considered initial magnetic
field $\beta(0)=5\times10^{-5}$ and a positive value for the initial shear component along the $z$ axis (the same
value as in the plot of $S+H$ in figure \ref{fig:G2}). The plots show that larger values for the magnetization
correspond to lower temperatures. The case with zero initial shear is displayed in figure \ref{fig:G4}.}
\end{center}
\end{figure}

 It is worthwhile commenting on the relation between the
temperature, magnetic field and magnetization in magnetized gases. In
an ``earth bound'' context when the gas is not self--gravitating, the
magnetic field acts as an ``external'' agent whose effect is to
increase the magnetization by aligning the magnetic moments of the
electrons. For higher temperatures the internal energy increases and
there is more resistance to this effect, resulting in a relatively low magnetization
with the electron magnetic moments alignment being relatively random, while
at lower temperatures the opposite effect occurs: magnetization is high and magnetic
moments strongly align with the magnetic field. However, for a self--gravitating gas the magnetic
field is no longer ``external'', but generated by the electron magnetic moments themselves, hence the relation
between these variables cannot be controlled: it emerges from the dynamics of the
gas and depends on the interplay of initial conditions. As we show in figure \ref{fig: disc1}, a large
magnetization coincides with a large magnetic field for relatively low temperatures.

It is important to recall that the study we have undertaken here is based on an equation of state
and a thermodynamical potential ($\Omega$) that come from a one-loop approximation.
The behavior of the system in the regime ($T^2\ll eB\ll m_e^2$) may seem strange:  the energy density does not
increase with the temperature
%                               CORRECCION ALAIN
%  EN NINGUN MOMENTO SE COMPARA CON NADA DEL ARTICULO \cite{Alain_e-} Las condiciones iniciales son diferentes
%and the resulting point--like singularity for low initial temperature is different from that obtained for the
%case $T=0$ (a cigar--like singularity) \cite{Alain_e-}.     
All this could follow as a consequence that the one--loop approximation may not be appropriate in some regimes
for a magnetized system in presence of finite density and temperature.
This issue deserves a separate study, possibly in the context of non-perturbative calculations at high magnetic
fields, which  explore a different approach that may be applicable to the magnetized gas that we have examined
here (see \cite{Ayala:2004dx} and references quoted therein).

\section{Conclusion}
We have examined the dynamical and thermodynamical behavior of a magnetized, self--gravitating
electron gas at finite temperature, taken as the source of a simplified Bianchi I space--time represented by a
Kasner metric, which is the simplest geometry that allows us identify the magnetic field as the main source of
anisotropy. We regard this configuration as a toy model that roughly approximates a grand canonical subsystem
of a magnetized electron source in the conditions prevailing near of the center and rotation axis of a compact
object (in which spatial gradients of physical and kinematic variables may be regarded as negligible).
The resulting Einstein--Maxwell field equations were transformed into a system of non--linear autonomous
evolution equations, which were solved numerically in the collapsing regime for a chemical potential:
$\mu=2m_{e}$, a magnetic field in the range $\hbox{B}\thicksim 10^{7}-10^{8}\hbox{G}$ and temperatures
$\hbox{T}\thicksim10^{3}-10^{7}\hbox{K}$.

For all initial conditions that we considered the gas evolves into a collapsing singularity, which can be
(depending on the initial conditions) isotropic (``point--like'') or anisotropic (``cigar--like''). We found
that for lower initial values of the magnetic field and temperatures ($B\backsim 10^{8}\hbox{G}, T\backsim
10^{3}\hbox{K}$) the resulting singularity is always point--like singularity, independently of the initial
values of the shear. This result may be connected with the stability of volume elements in less idealized
configurations, an issue that is outside the scope of this paper and deserves a proper examination elsewhere.
%                                        CORRECCION  ALAIN
% NO ESTOY DE ACUERDO CON ESTO EN EL ARTICULO \cite{Alain_e-} las condiciones iniciales son diferentes por tanto no 
%se puede comparar con el, a menos q Ismael haga hecho esta comparacion, pero pienso no la hizo.
%
%Also, the effect of the initial shear on the resulting type of singularity is qualitatively different from
%that obtained for the case $T=0$ that we examined in \cite{Alain_e-}, in which an initial positive value
%of the shear eigenvalue in the direction of the magnetic field always lead to a cigar--like singularity.
%

The collapse time decreases as the initial magnetic field increases, but we did not find a direct
proportionality relation between this time and the initial temperature (this may be a consequence of
having assumed a high density regime). The behavior of the temperature as the collapse proceeds also
depends on initial conditions: for initial temperatures $\hbox{T}\thicksim10^{3}-10^{6}\hbox{K}$ the
gas cools down and for values $\hbox{T}\gtrsim10^{7}\hbox{K}$ the temperature increases.

Using the numerical solutions of the system we found a monotonic relation between the magnetization
and the magnetic field and between the energy density and the temperature for high temperature values,
but relation does not occur for low temperature values. This difference in behavior can be explained
by the predominance of the magnetic effects at low temperatures, though it may also be due to the
limitations introduced  by assuming an equation of state and a thermodynamical potential based on a
one--loop approximation. Looking at this issue is beyond the scope of this paper and will be pursued
in a separate work.

The study we have presented can be readily applied to examine hadronic systems (complying with suitable
balance conditions and adequate chemical potentials). The methodology we have used can also serve as
starting point to study the origin and the dynamics of primordial  cosmological magnetic fields. These
potential extensions of the present work are already under  consideration for future articles.

\section{Acknowledgements}
%\acknowledgements
{The work of A.P.M, A.U.R and I.D has been supported by \emph{Ministerio
de Ciencia, Tecnolog\'{\i}a y Medio Ambiente} under the grant CB0407
and the ICTP Office of External Activities through NET-35. APM acknowledges
to Prof R. Ruffini for his hospitality and financial support at International
Center for Relativistic Astrophysics Network-ICRANET. A.P.M.
also acknowledges the Program of Associateship TWAS-UNESCO-CNPq as
well as the hospitality and support of CBPF through the program
PCI-MCT. R.A.S. and A.U.R. acknowledge support from the research
grant \emph{SEP--CONACYT--132132}, and the TWAS-CONACYT fellowships.}

\begin{appendix}
\section{Transformation of the integrals}

The definitions of $\Gamma_{\parallel}$, $\Gamma_{\perp}$, $\Gamma_{U}$,
$\Gamma_{\eta}$ are given by equations (\ref{eq:Pxx})-(\ref{eq:Es etha}),
together with the coefficients (\ref{eq:C1})-(\ref{eq:C4}). Hence, their derivatives take the following form:
\begin{eqnarray}
\fl \Gamma_{U,\beta} &=& \left[\frac{1}{2}C_{3}\left(\mu,\phi\right)+\sum_{n=1}^{\infty}a{}_{n}^{2}C_{3}\left(\frac{\mu}{a_{n}},\frac{\phi}{a_{n}}\right)\right]\nonumber
\\
\fl &{}& +\beta\sum_{n=1}^{\infty}n\left[2C_{3}\left(\frac{\mu}{a_{n}},\frac{\phi}{a_{n}}\right)-\frac{a_{n}}{\phi}C_{6}\left(\frac{\mu}{a_{n}},\frac{\phi}{a_{n}}\right)\right],
\\
\fl \Gamma_{U,\phi} &=& \frac{\beta}{\phi^{2}}\left\{ \frac{1}{2}\left[C_{6}\left(\mu,\phi\right)-\mu C_{5}\left(\mu,\phi\right)\right]+\sum_{n=1}^{\infty}a_{n}^{2}\left[a_{n}C_{6}\left(\frac{\mu}{a_{n}},\frac{\phi}{a_{n}}\right)-\mu C_{5}\left(\frac{\mu}{a_{n}},\frac{\phi}{a_{n}}\right)\right]\right\},\nonumber  \\
\fl &{}&\\
\fl \Gamma_{U,\mu} &=& \frac{\beta}{\phi}\left\{ \frac{1}{2}C_{5}\left(\mu,\phi\right)+\sum_{n=1}^{\infty}a_{n}^{2}C_{5}\left(\frac{\mu}{a_{n}},\frac{\phi}{a_{n}}\right)\right\},
\\
\fl \Gamma_{\eta,\beta} &=& \frac{\beta}{\phi}\left[\frac{1}{2}C_{4}\left(\mu,\phi\right)+\sum_{n=1}^{\infty}a{}_{n}C_{4}\left(\frac{\mu}{a_{n}},\frac{\phi}{a_{n}}\right)\right]\nonumber
\\
\fl &{}& \qquad\qquad\qquad+\beta\sum_{n=1}^{\infty}n\left[\frac{1}{a_{n}}C_{4}\left(\frac{\mu}{a_{n}},\frac{\phi}{a_{n}}\right)-\frac{1}{\phi}C_{5}\left(\frac{\mu}{a_{n}},\frac{\phi}{a_{n}}\right)\right],
\\
\fl \Gamma_{\eta,\phi} &=& \frac{\beta}{\phi^{2}}\left\{ \frac{1}{2}\left[C_{5}\left(\mu,\phi\right)-\mu C_{7}\left(\mu,\phi\right)\right]+\sum_{n=1}^{\infty}a_{n}\left[a_{n}C_{5}\left(\frac{\mu}{a_{n}},\frac{\phi}{a_{n}}\right)-\mu C_{7}\left(\frac{\mu}{a_{n}},\frac{\phi}{a_{n}}\right)\right]\right\},\nonumber
\\
\fl&{}&\\
\fl \Gamma_{\eta,\mu} &=& \frac{\beta}{\phi}\left\{ \frac{1}{2}C_{7}\left(\mu,\phi\right)+\sum_{n=1}^{\infty}a_{n}C_{7}\left(\frac{\mu}{a_{n}},\frac{\phi}{a_{n}}\right)\right\},
\end{eqnarray}
where the coefficients $C_{5},\: C_{6},\: C_{7}$ are given by:
\begin{eqnarray}
C_{5}\left(\mu,\phi\right)=\intop_{0}^{\infty}\frac{h\left(\frac{\sqrt{1+x^{2}}-\mu}{\phi}\right)}{\sqrt{1+x^{2}}}dx,\label{eq:C5}
\\
C_{6}\left(\mu,\phi\right)=\intop_{0}^{\infty}\frac{x^{2}}{\sqrt{1+x^{2}}}h\left(\frac{\sqrt{1+x^{2}}-\mu}{\phi}\right)dx,
\\
C_{7}\left(\mu,\phi\right)=\intop_{0}^{\infty}h\left(\frac{\sqrt{1+x^{2}}-\mu}{\phi}\right)dx,\label{eq:C7}
\end{eqnarray}
while $h\left(x\right)=1/\left(4\cosh^{2}\left(x/2\right)\right)$. The integrals (\ref{eq:C1})-(\ref{eq:C4}), (\ref{eq:C5})-(\ref{eq:C7})
take the form:
\begin{equation}
C_{i}\left(\phi,\mu\right)=\Bigg\{
%\begin{cases}
  \begin{array}{lr}
\intop_{0}^{\infty}f_{i}\left(x\right)\frac{1}{1+\exp\left(\frac{\sqrt{1+x^{2}}-\mu}{\phi}\right)}\,dx\,, &
i=\overline{1,4},
\\
\intop_{0}^{\infty}f_{i}\left(x\right)h\left(\frac{\sqrt{1+x^{2}}-\mu}{\phi}\right)dx, &
i=\overline{5,7}. \\
  \end{array}
%\end{cases}
\end{equation}
Integrating by parts and making a change of variables $\nu=[\sqrt{1+x^{2}}-\mu]/\phi$
we finally obtain:
\begin{equation}
\fl
\qquad \qquad \; C_{i}\left(\phi,\mu\right)=\Bigg\{
\begin{array}{lr}
\intop_{-\frac{\mu-1}{\phi}}^{\infty}F_{i}\left(\sqrt{\left(\phi\,
\nu+\mu\right)^{2}-1}\right)h\left(\nu\right)d\nu\,, & i=\overline{1,4},
\\
\intop_{-\frac{\mu-1}{\phi}}^{\infty}F_{i}\left(\sqrt{
\left(\phi\,\nu+\mu\right)^{2}-1}\right)H\left(\nu\right)d\nu\,, & i=\overline{5,7}, \\
\end{array}
\end{equation}
with
\begin{equation}
\fl
\quad \begin{array}{ccc}
F_{i}\left(x\right)=\intop_{0}^{x}f\left(y\right)dy,
& h\left(\nu\right)=\left[4\cosh^{2}\left(\nu/2\right)\right]^{-1},
& H\left(\nu\right)=\frac{e^{\nu}\left(e^{\nu}+1\right)}{\left(e^{\nu}+1\right)^{3}}. \\
\end{array}
\end{equation}
\end{appendix}

\end{document}